\documentclass[twocolumn,prd, superscriptaddress, showpacs]{revtex4-1}
\usepackage{hyperref,xcolor}
\usepackage{amsmath,amssymb}
\usepackage{graphicx}
\usepackage{xcolor}
\usepackage{bm}
\usepackage{textcomp}

\newcommand{\rd}{\ensuremath{\mathrm{d}}}

\providecommand{\abs}[1]{\lvert#1\rvert}

\newcommand{\expv}[1]{\left\langle#1\right\rangle}

\newcommand{\bphi}{\bm{\Phi}}
\newcommand{\bdphi}{\bm{\delta\Phi}}
\newcommand{\be}{\begin{equation}}
\newcommand{\ee}{\end{equation}}
\newcommand{\benn}{\nonumber\begin{equation}}
\newcommand{\eenn}{\nonumber\end{equation}}
\def\bea{\begin{eqnarray}} \def\eea{\end{eqnarray}}
\def\beann{\begin{eqnarray*}} \def\eeann{\end{eqnarray*}}
\def\lsim{\raise0.3ex\hbox{$<$\kern-0.75em\raise-1.1ex\hbox{$\sim$}}}
\def\gsim{\raise0.3ex\hbox{$>$\kern-0.75em\raise-1.1ex\hbox{$\sim$}}}

\renewcommand{\Re}{{\rm Re}}
\renewcommand{\Im}{{\rm Im}}

\begin{document}
{\hfill \texttt{\footnotesize CERN-PH-TH/2014-091}}
\vfill
\title{Extended mean field study of complex $\varphi^4$-theory at finite density and temperature}
\date{\today}
\author{Oscar Akerlund}
\affiliation{Institut f\"ur Theoretische Physik, ETH Z\"urich, CH-8093 Z\"urich, Switzerland}
\author{Philippe de Forcrand}
\affiliation{Institut f\"ur Theoretische Physik, ETH Z\"urich, CH-8093 Z\"urich, Switzerland}
\affiliation{Physics Department, CERN, TH Unit, CH-1211 Geneva 23, Switzerland}
\author{Antoine Georges}
\affiliation{Centre de Physique Th\'eorique, CNRS, \'Ecole Polytechnique, 91128 Palaiseau, France}
\affiliation{Coll\`ege de France, 11 place Marcelin Berthelot, 75005 Paris, France}
\affiliation{DPMC, Universit\'e de Gen\`eve, 24 quai Ernest Ansermet, CH-1211 Geneva, Switzerland}
\author{Philipp Werner}
\affiliation{Department of Physics, University of Fribourg, 1700 Fribourg, Switzerland}

\begin{abstract}
We review the extended mean field theory (EMFT) approximation and apply it to complex, scalar $\varphi^4$ theory
on the lattice. We study the critical properties of the Bose condensation driven by a nonzero chemical
potential $\mu$ at both zero and nonzero temperature and determine the $(T,\mu)$ phase diagram. The results are in very 
good agreement with recent Monte Carlo data for all parameter values considered.
EMFT can be formulated directly in the thermodynamic limit which allows us to study lattice spacings for which Monte
Carlo studies are not feasible with present techniques. We find that the EMFT approximation accurately reproduces 
many known phenomena of the exact solution, like the ``Silver Blaze'' behavior at zero temperature and dimensional reduction
at finite temperature.
\end{abstract}

\maketitle

\section{Introduction}\label{sec:introduction}\noindent
One serious obstacle in lattice field theory and computational physics is the so-called ``sign problem,'' 
which spoils the probabilistic interpretation of the partition function and thus a foundation of the
otherwise powerful Monte Carlo method. A sign (or phase) problem may have different origins.
On the one hand, the statistics of the fields might cause some configurations to appear with a negative (fermions) or
complex (anyons) weight. While it is possible, in principle, to consider suitable subsets of the configuration
space~\cite{Bloch:2013} or to use another set of variables~\cite{Chandrasekharan:2013rpa} to end up with only non-negative weights,
appropriate subsets or new variables have only been found for a small number of models so far.  
On the other hand, the action itself can be complex leading to sign problems even in bosonic systems.
A typical example for this case is when a chemical potential is introduced, which creates an asymmetry between 
particles and antiparticles. Also here, the sign problem can sometimes be solved by considering a different
set of variables, like in the world-line Monte Carlo approach~\cite{Endres:2006xu, Gattringer:2012df}.
Recent progress in the understanding of the complex Langevin equations~\cite{Aarts:2009,Aarts:2013uxa} and
gauge cooling~\cite{Seiler:2012wz} has promoted yet another approach for simulating models with complex actions.

Mean field (MF) methods, although approximative, can be useful alternatives. They are computationally cheap, and
many results can be obtained analytically or at least semi-analytically. Furthermore, most of the time the
symmetries of the Lagrangian can be used to make the action real, hence avoiding the sign problem.
Standard mean field methods have however some obvious shortcomings. Although mean field theory is known to
reproduce the correct qualitative critical behavior at and above the upper critical dimension (up to logarithmic corrections),
quantitative predictions are usually very approximative. Another shortcoming of mean field theory is that it cannot be used to
determine correlation functions or to study nonzero temperature. A simple extension of mean field theory
which aims to overcome these limitations is EMFT \cite{Pankov:2002, Akerlund:2013}, which incorporates
self-consistency at the level of the propagator.

In this paper we review the derivation of the EMFT equations and apply them to complex scalar $\varphi^4$ theory, one
of the simplest models with a sign problem of the second type described above. The chemical potential couples
to a conserved charge which is a consequence of Noether's theorem and of the global U(1) symmetry. The model describes
a relativistic Bose gas and its properties are well studied. It is one of the models where the world-line
formulation \cite{Gattringer:2012df} can be applied and we will take advantage of this to evaluate the quality of
the approximation.

The rest of the paper is structured as follows: In Sec.~\ref{sec:phi4} we will briefly introduce the
studied model before going on to presenting the mean field and EMFT formulations in Secs.~\ref{sec:mf}~and~
\ref{sec:emft}, respectively. In Sec.~\ref{sec:results} we present our results, and Sec. \ref{sec:conc}
is devoted to a discussion of our conclusions.

\section{$\varphi^4$ Theory}\label{sec:phi4}

\noindent
$\varphi^4$ theories are important quantum field theories in many respects. Even the simplest incarnation, with a 
single real scalar field, exhibits interesting phenomena like spontaneous symmetry breaking with a second-order
phase transition. The U(1) symmetric complex $\varphi^4$ theory with nonzero chemical potential is one of the
simplest models which has a sign problem. One important application of the latter is in the Standard Model Higgs
sector, which consists of a two-component complex $\varphi^4$ theory.
 
In dimensions higher than two, complex $\varphi^4$ exhibits a second-order phase transition as a function of the
chemical potential, $\mu$. At low $\mu$ the system is a dilute Bose gas which Bose condenses above a critical
chemical potential, $\mu_c$. We are mainly interested in the four-dimensional case but for the sake of generality
we will work in $d$-dimensions and specify $d$ only when necessary. The Lagrangian density of complex scalar
$\varphi^4$ theory at finite chemical potential reads
\begin{align}
\mathcal{L}[\varphi(x)] &= \partial_\nu \varphi^*(x)\partial^\nu \varphi(x) - \left(m_0^2-\mu^2\right)\abs{\varphi(x)}^2\nonumber\\
&\phantom{=} - \lambda\abs{\varphi(x)}^4 +i \mu j_0(x),\\
j_\nu(x) &= \varphi^*(x)\partial_\nu\varphi(x) - \partial_\nu\varphi^*(x)\varphi(x),
\end{align}
using a $d$-dimensional Minkowski metric, $(+,-,\ldots,-)$. $\left(m_0^2-\mu^2\right)$ is a physically irrelevant shift of the
bare mass, which is convenient when putting the theory on the lattice, where the $\mu^2$ will drop out. 
$j_\nu$ is the conserved current due to the global continuous $U(1)$ symmetry, $\varphi(x) \leftrightarrow e^{i\theta}\varphi(x) ~\forall x$, with the conserved charge 
\be
Q = i\int\rd^{d-1}x\;j_0(x).
\ee
The charge represents the number of particles minus the number of antiparticles and a positive $\mu$ thus favors the creation of particles over antiparticles
and renders the Lagrangian density (and action) complex.

After Wick rotating time to the imaginary axis to obtain a Euclidean metric, we discretize the action and put it on a
regular $d$-dimensional hypercubic lattice with lattice spacing $a$. The chemical potential is associated with the (imaginary) time 
direction which will be referred to as $t$. All parameters are understood to be in terms of the lattice spacing, so we
refrain from explicitly writing for example $a\mu$ instead of $\mu$ without causing confusion.
With $\eta\equiv m_0^2 + 2d$ we arrive at the usual lattice action
\begin{align}\label{eq:action_latt}
S &= \sum_x\left(\eta\abs{\varphi_x}^2 + \lambda\abs{\varphi_x}^4 \phantom{\sum_\nu^4}\right.\nonumber\\
&\phantom{=\sum_x}\left.- \sum_{\nu=1}^d\left[e^{-\mu\delta_{\nu,t}}\varphi^*_x\varphi_{x+\hat{\nu}} + e^{\mu\delta_{\nu,t}}\varphi^*_x\varphi_{x-\hat{\nu}}\right]\right).
\end{align}
Because of different couplings in the forward and backward time direction the action is complex when $\mu\neq0$.
This prevents the usual probabilistic interpretation of the partition function and Monte Carlo methods cannot be
blindly applied. The sign problem can be circumvented by a change of variables which allows to express the action in terms
of world lines. The partition function can then be sampled using a worm algorithm,
see e.g.~\cite{Endres:2006xu,Gattringer:2012df}. Another alternative is to use a complex Langevin method \cite{Aarts:2009,Aarts:2009hn}.
We will consider a mean field like approximation and thus also avoid the sign problem.

\section{Mean Field Theory}\label{sec:mf}\noindent
The upper critical dimension of the complex $\varphi^4$ theory is $d_\text{uc} = 4$, so we expect that the mean field solution
will show a qualitatively correct behavior and provide a first approximation to quantitative results. Taking the action
in Eq.~\eqref{eq:action_latt} and setting the field to its expectation value (``Weiss field''), $\varphi_x = \expv{\varphi}$,
for all $x\neq0$, gives us the single-site mean field action
\be
S_\text{MF} = \eta\abs{\varphi_0}^2 +\lambda\abs{\varphi_0}^4- 4\expv{\varphi} \Re[\varphi_0](d-1+\cosh(\mu)).
\ee
We have used the $U(1)$ symmetry to rotate the expectation value to the real axis. The magnitude of the expectation
value, $\expv{\varphi}$, is determined self-consistently by requiring
\be
\expv{\varphi_0}_{S_\text{MF}} = \expv{\varphi}.
\ee

It is easy to check that there is a second-order phase transition at a critical chemical potential, $\mu_c$,
whose exact value depends on $d,\eta$ and $\lambda$. By expanding $\exp(-S_\text{MF})$ in powers of $\expv{\varphi}$,
demanding self-consistency for $\expv{\varphi}$ and letting it go to zero, we find the critical chemical potential:
\be
\cosh\mu_c(\eta,\lambda) =\frac{\sqrt{\lambda}}{2\frac{\exp\left(-K^2\right)}{\sqrt{\pi}\text{Erfc}\left(K\right)}-2K} +(1-d),
\ee
with $K = \tfrac{\eta}{2\sqrt{\lambda}}$. We can determine the continuum limit in the mean field approximation by
searching the critical value of $\eta$ for which $\mu_c$ vanishes. For $d=4$ and $\lambda=1$ we find $\eta_c=7.51366$.

In order to improve on standard mean field theory, we would like to take also quadratic fluctuations into account.
To this end we apply EMFT, which self-consistently determines the local, or zero separation ($k$-integrated) Green's function, 
$G(\vec{r}=\vec{0},t=0)\equiv G_{xx}$.

\section{Extended Mean Field Theory}\label{sec:emft}

\subsection{Formalism}
\noindent
EMFT \cite{Akerlund:2013} is based on the work of Pankov {\it et al.} \cite{Pankov:2002} and is a systematic extension of standard
mean field theory in which all fluctuations up to a given order in the field can be taken into account. Just as standard mean
field theory, EMFT is a single site approximation in which the fields around a single site are treated as an effective
bath which is self-consistently determined. Upon integrating out the effective bath an infinite series of
self-interactions is generated and the truncation of this series determines the level of self-consistency. For each
individual term the full local interaction is taken into account: there is no expansion in the coupling and the
method is not restricted to weak couplings. We have previously applied the method to the real scalar $\varphi^4$ theory
and obtained very good results \cite{Akerlund:2013}.

EMFT can also be viewed as the local time limit of dynamical mean field theory (DMFT), which is extensively used in the
condensed matter community, see e.g. \cite{Georges:1996} for a review. In DMFT the effective Weiss field is a function of one
coordinate, usually ``time'' (hence the name). The effective model is thus a world line frozen in space with the full local interaction plus
nonlocal interactions along the world line. In DMFT these nonlocal interactions are almost always truncated at the quadratic terms, which
implies that the free effective theory is exactly solvable and the effective field can be self-consistently 
determined by identifying the Green's function with an approximation of the local Green's function of the full theory.
In the local time limit, i.e. EMFT, the world line is just one point and the effective fields can be thought of as
coupling constants in a polynomial potential. These coupling constants can in principle be self-consistently determined
by matching local $n$-point correlators to moments of the effective one-site model.

We will now derive the EMFT effective action and the self-consistency equations. For convenience we will use
a slightly unconventional notation intended to  make the derivation more transparent. The action,
Eq.~\eqref{eq:action_latt}, in this notation reads
\begin{equation}
S = \displaystyle\sum_x\left[\!-\sum_\nu \bphi^\dagger_{x +\widehat{\nu}}\bm{E}(\mu\delta_{\nu,t})\bphi_x \!+\! \frac{\eta}{2}\abs{\bphi_x}^2 \!+\! \frac{\lambda}{4}\abs{\bphi_x}^4\!\right], 
\end{equation}
with 
\begin{equation}
\bphi^\dagger = (\varphi^*,\varphi),\quad \bm{E}(x) = \begin{pmatrix} e^{-x} & 0 \\ 0 & e^{x} \end{pmatrix}.
\end{equation}
In the free case ($\lambda=0$) the action is quadratic in $\bphi$ and the inverse of the connected Green's function in Fourier space can be easily expressed as a matrix,
\begin{align}
&\widetilde{\bm{G}}_0^{-1}(k) = \expv{\bphi\bphi^\dagger}_c =  \expv{\bphi\bphi^\dagger} - \expv{\bphi}\expv{\bphi}^\dagger\nonumber\\ 
= &\begin{pmatrix}\eta - 2\displaystyle\sum_{\nu=1}^d\cos\left(k_\nu - i\mu\delta_{\nu,t}\right) & 0 \\
  0 & \eta - 2\displaystyle\sum_{\nu=1}^d\cos\left(k_\nu  +i\mu\delta_{\nu,t}\right)\end{pmatrix}.\label{eq:free_gf}
\end{align}
(We put a tilde on Fourier transformed quantities.) 
The full lattice Green's function can then be expressed as
\begin{equation}\label{eq:gf_latt}
\widetilde{\bm{G}}^{-1}(k) = \widetilde{\bm{G}}_0^{-1}(k) - \widetilde{\bm{\Sigma}}(k),
\end{equation}
where $\bm{\Sigma}$ is the self-energy due to $\lambda\neq0$. This point is paramount to EMFT and similar methods. The 
Green's function is known at some point in parameter space, at $\lambda=0$ in this case, and the deviation of the full
Green's function from the known one can be quantified by a function that depends on the interaction, $\lambda$.
The aim is then to find a simpler but (at least approximately) equivalent model which can be solved more easily
than the full model. If the simpler model yields the same interaction-dependent deviation of the Green's function as
the full model, solving the simpler model is equivalent to solving the full model. If the simpler model is only
approximately equivalent then naturally an approximate solution is obtained. It can also happen that the simpler
model is a valid approximation only in some limited regime such that it can only be used to determine some subset of
all observables of the full model. We will now derive an equivalent model to Eq.~\eqref{eq:action_latt} which
will turn out to be valid for local observables.

As in any mean field approach we expand the field $\bphi$ around its (real) mean,
$\expv{\bphi} = \bar{\bm{\phi}}$: $\bphi = \bar{\bm{\phi}} + \bdphi$. Focusing on the field at the origin,
$\bphi_0$, the action can be written as
\begin{align}
S &= S_0 + \delta S + S_\text{ext},\nonumber\\
S_0 &= \frac{\eta}{2}\abs{\bphi_0}^2 + \frac{\lambda}{4}\abs{\bphi_0}^4 - 2\bar{\bm{\phi}}^\intercal\bphi_0(d-1+\cosh(\mu)),\nonumber\\ 
\delta S &= -\sum_{\pm\nu}\bdphi_{0+\hat{\nu}}^\dagger\bm{E}(\pm\mu\delta_{\nu,t})\bdphi_0.\label{eq:action_onesite}
\end{align}
The term $S_\text{ext}$ does not depend on $\varphi_0$ and is irrelevant for our purpose. The term
$\delta S$ contains the interaction of $\bphi_0$ 
with its nearest
neighbors $\bphi_{0\pm\hat{\nu}}$, which are to be integrated out. The field at those sites is collectively
denoted by $\varphi_\text{ext}$. The integration over $\varphi_\text{ext}$ is formally done by replacing $\delta S$ by
its cumulant expansion with respect to $S_\text{ext}$, 
\begin{equation}
Z = \!\!\int\!\!\rd\varphi_0\mathcal{D}\varphi_\text{ext}\,e^{-S_0-\delta S - S_\text{ext}}=\!\int\!\!\rd\varphi_0\,e^{-S_0-\expv{\delta S}_\text{ext}^\text{C}},\label{eq:part_function}
\end{equation}
where $\expv{\delta S}_\text{ext}^\text{C}$ denotes the cumulant expansion. To second order in the fluctuation $\bdphi_0$
it reads:
\begin{widetext}
\begin{equation}
\expv{\delta S}_\text{ext}^\text{C}\approx\expv{\sum_{\pm\nu}\bdphi_{\hat{\nu}}^\dagger\bm{E}(\pm\mu\delta_{\nu,t})\bdphi_0}_{\!\!\!S_\text{ext}} \!\!\!\!\!\!+ \frac{1}{2}\expv{\sum_{\pm\nu}\bdphi_{\hat{\nu}}^\dagger\bm{E}(\pm\mu\delta_{\nu,t})\bdphi_0 \sum_{\pm\rho}\bdphi_{\hat{\rho}}^\dagger\bm{E}(\pm\mu\delta_{\rho,t})\bdphi_0}_{\!\!\!S_\text{ext}} \!\!\!\!\!= 0 + \frac{1}{2}\bdphi_0^\dagger\bm{\Delta}\bdphi_0.\label{eq:cum_exp}
\end{equation}
\end{widetext}
The first term is zero because $\expv{\bdphi_{\hat{\nu}}}_{S_\text{ext}}=0$ by definition and $\bm{\Delta}$ is an unknown real,
symmetric matrix which is related to the second term and will be determined self-consistently. $\bm{\Delta}$ is given by a sum
of real bosonic propagators and is therefore real. It is symmetric since the fields commute, i.e.  
$\expv{\varphi_i\varphi_j}=\expv{\varphi_j\varphi_i}$. In our case we can parametrize $\bm{\Delta}$ as 
\begin{equation}
  \bm{\Delta} =
  \begin{pmatrix}
    \Delta_{11} & \Delta_{12}\\\Delta_{12} & \Delta_{11}
  \end{pmatrix}.
\end{equation}

We truncate the cumulant expansion at quadratic order in $\bdphi$ for simplicity. In principle, keeping-higher order
terms provides a way to systematically improve the approximation but it may be hard to find suitable self-consistency
conditions for the higher-order couplings. Inserting the truncated expansion in
Eq.~\eqref{eq:part_function} and using $\bdphi_0 = \bphi_0-\bar{\bm{\phi}}$ yields an effective one-site action
\begin{align}\label{eq:action_emft}
S_\text{EMFT} &= \frac{1}{2}\bphi^\dagger\left(\eta\bm{I} - \bm{\Delta}\right)\bphi + \frac{\lambda}{4}\abs{\bphi}^4 \\
&\phantom{=}- 2\phi\Re[\varphi](2(d-1+\cosh(\mu))-\Delta_{11}-\Delta_{12}),\nonumber
\end{align}
which can effortlessly be solved.

Like the full Green's function above, the EMFT Green's function can be expressed as a free part and a self-energy,
\be\label{eq:gf_emft}
\bm{G}^{-1}_\text{EMFT} = \eta\bm{I} - \bm{\Delta} - \bm{\Sigma}_\text{EMFT}.
\ee
Replacing the full self-energy in Eq.~\eqref{eq:gf_latt} by the EMFT self-energy completes the mapping.  
It should 
now be noted that since the effective EMFT model is a single site model, we can only expect it to correctly reproduce
local observables. 
(If we had taken the entire cumulant expansion in
Eq.~\eqref{eq:cum_exp} then the effective action would exactly correspond to the full theory
and would generate all local observables.) Substituting $\bm{\Sigma}_\text{EMFT}$ into Eq.~\eqref{eq:gf_latt} yields 
\be\label{eq:gf_sub}
\widetilde{\bm{G}}^{-1}(k) \approx \bm{G}^{-1}_\text{EMFT} + \bm{\Delta} -2\sum_{\nu=1}^d\cos\left(k_\nu - i\mu\delta_{\nu,t}\right)\bm{I}.
\ee
Notice that we here have neglected that the imaginary part of the two diagonal elements in
$\widetilde{\bm{G}}$ differ. On the one hand this is justified since after integrating over all $k$ the
result will be real. On the other hand it allows us to easily invert the propagator and one can show
that the neglected terms in $\widetilde{\bm{G}}(k)$ are regular as $k_t$ goes to zero whereas the propagator
itself diverges at the critical point, so this approximation will at most change the UV behavior of the theory.

In order to fix $\bm{\Delta}$ we need to identify the local full
lattice Green's function with the EMFT Green's function, which together with the self-consistency for
$\phi$ yields a set of three coupled self-consistency equations,
\begin{align}
\phi &= \expv{\varphi}_{S_\text{EMFT}},\\
\int\frac{\rd^dk}{(2\pi)^d}\widetilde{\bm{G}}(k) &\equiv \bm{G}_{xx} = \bm{G}_\text{EMFT},\label{eq:sc_gf}
\end{align}
where the matrix equation~\eqref{eq:sc_gf} yields two independent equations, one for the diagonal element and one
for the off-diagonal element. These equations are satisfied at stationary points of the (approximate) local free energy
functional~\cite{Potthoff:2003}.

In order not to be hampered by high dimensionality and/or many components in the field it is important to evaluate the $k$ integral in an
efficient way. By diagonalizing $\widetilde{\bm{G}}^{-1}(k)$ we can transform the $d$-dimensional integral into a one-dimensional one, which gives 
\begin{align}
\bm{G}_{xx}^\pm &= \frac{1}{2}\int_0^\infty\!\!\!\!\!\rd\tau\,\left(e^{-\tau\left(\frac{1}{2\langle(\Re\varphi)^2\rangle}+\Delta_{11}+\Delta_{12}\right)}\right.\\
&\phantom{= \frac{1}{2}\int_0^\infty\!\!\!\rd\tau\,(}\left.\pm e^{-\tau\left(\frac{1}{2\langle(\Im\varphi)^2\rangle}+\Delta_{11}-\Delta_{12}\right)}\right)(I_0(2\tau))^d,\nonumber
\end{align}
where $\bm{G}_{xx}^+$ is the diagonal element, $\bm{G}_{xx}^-$ the off-diagonal element and $I_0(x)$ is the zeroth modified Bessel function of the first kind.
More details on the transformation of the integral can be found in Appendix~\ref{app:kint}.

\subsection{Finite lattices}

\noindent
Because the self-consistency equation~\eqref{eq:sc_gf} involves a $k$ sum, 
 the results will depend on how we define our lattice model.
For example, we can treat nonzero temperature simply by summing over a finite number of timelike momenta, $k_t$.
We can equally well consider a finite sized spatial box. In fact, we can easily study the model on any hypercubic lattice with 
$(N_x,N_y,N_z,N_t)\in\{2,\ldots,\infty\}$.

\subsection{Observables}\noindent
Through the self-consistency equations we have direct access to the expectation value of the field and the local Green's function.
Another interesting and nontrivial observable is the density, $n$, which is defined as the partial derivative of the free energy,
or logarithm of the partition function, with respect to the chemical potential. By recasting the nearest neighbor interaction of
the original action, Eq.~\eqref{eq:action_latt}, in Fourier space one finds that the density can be expressed as 
\begin{widetext}
\begin{equation}\label{eq:dens}
n =2\sinh\mu\expv{\varphi}^2 \!+ 2\left(\sinh\mu\!\int\!\!\frac{\rd^d k}{(2\pi)^d}\Re[\expv{\varphi^*(k)\varphi(k)}_c]\cos(k_t)-\cosh\mu\!\int\!\!\frac{\rd^d k}{(2\pi)^d}\Im[\expv{\varphi^*(k)\varphi(k)}_c]\sin(k_t)\right).
\end{equation}
\end{widetext}
The correlator $\expv{\varphi^*(k)\varphi(k)}_c$ is nothing else than the connected Green's function, which in our local approximation
is given by the diagonal elements of $\widetilde{\bm{G}}({k})$. In Appendix~\ref{app:dens} we show that the two weighted integrals
cancel at zero temperature and exhibit a weak $\mu$ dependence at nonzero temperatures.
This is exactly the (pseudo) Silver Blaze behavior \cite{Cohen:2003}.

\subsection{Extra constraints}

\noindent
We have seen in Eq.~\eqref{eq:gf_sub} that EMFT produces an approximation of the full Green's function. It is therefore tempting
to extract observables from it, the prime example being the masses of $\varphi_{1,2}$, $m^2_i = \bm{G}_{ii}^{-1}(0)$. This is also fine
as long as one keeps in mind that the resulting masses are only approximate. In particular one may obtain a nonzero mass for
$\varphi_2$ although the Nambu-Goldstone theorem tells us it must be zero. That this may happen can be quite easily demonstrated.
Consider the (exact) local propagator,
\begin{align}
  G_{xx} &= \int \rd k\, (\widetilde{G}_0(k)^{-1} - \widetilde{\Sigma}(k))^{-1} \\
  &\equiv Z^{-1}\int \rd k\, (M^2_\text{exact} + \hat{k}^2 + \widehat{\widetilde{\Sigma}}(k))^{-1},\nonumber
\end{align}
where $\hat{k}$ is the lattice momentum and $Z$ is the wave function renormalization. This is matched to the EMFT local propagator
through the self-consistency equation~\eqref{eq:sc_gf},
\begin{align}
  G_\text{EMFT} &= \int \rd k\, (\widetilde{G}_0(k)^{-1} - \Sigma_\text{EMFT})^{-1} \\
  &\equiv \int \rd k\, (M^2_\text{EMFT} + \hat{k}^2)^{-1}.\nonumber
\end{align}
Comparing the two equations above we see that $M_\text{exact}^2$ and $M_\text{EMFT}^2$ do not have to coincide for the local Green's functions to be equal. 
Thus, whilst $ZM_\text{exact}^2$ is the curvature of the effective potential and has a zero eigenvalue, the same
need not apply to $M_\text{EMFT}^2$. This argument is independent of whether we truncate the cumulant expansion or not. We can of course
explicitly calculate the effective potential, which by construction respects the U(1) symmetry and correctly has a flat direction at
its minimum. Another option is to slightly modify the EMFT equations to force the Goldstone mode to be massless by introducing an extra constraint.

To do so we first extract the mass matrix $\bm{M}^2$ from Eq.~\eqref{eq:gf_sub}:
\begin{align}\label{eq:mass_matrix}
  \widetilde{\bm{G}}^{-1}(\vec{0},k_t) &\propto \bm{M}^2 + k_t^2\bm{I} \\
  & = \frac{\bm{G}^{-1}_\text{EMFT} + \bm{\Delta}}{\cosh\mu}-\frac{2(d+\cosh\mu-1)}{\cosh\mu}\bm{I} + k_t^2\bm{I}.\nonumber  
\end{align}
As we have seen above, there is no guarantee that there will be massless mass eigenstates at the self-consistent fixed point.
We will enforce this by hand with an additional parameter. It is a fact that the momentum dependence of the interacting
Green's function differs from that of the free Green's function so it is natural to introduce the new parameter in such a way that
the momentum dependence is changed. Consider the substitution,  
\begin{equation}
  \bm{\Sigma}(k) \to \bm{\Sigma}_\text{EMFT}, 
\end{equation}
that we made in Eq.~\eqref{eq:gf_latt} to obtain Eq.~\eqref{eq:gf_sub} via Eq.~\eqref{eq:gf_emft}.
We now propose the alternative substitution
\begin{equation}
  \bm{\Sigma}(k) \to \bm{\Sigma}_\text{EMFT} + 2(Z-1)\sum_{\nu=1}^d\cos(k_\nu-i\mu\delta_{\nu,t})\bm{I}
\end{equation}
which leads to
\be\label{eq:gf_sub2}
\widetilde{\bm{G}}^{-1}(k) \approx \bm{G}^{-1}_\text{EMFT} + \bm{\Delta} -2Z\sum_{\nu=1}^d\cos\left(k_\nu - i\mu\delta_{\nu,t}\right)\bm{I}
\ee
and the mass matrix
\begin{equation}
  \label{eq:mass_matrix2}
  \bm{M}^2 =  \frac{\bm{G}^{-1}_\text{EMFT} + \bm{\Delta}}{Z\cosh\mu}-\frac{2(d+\cosh\mu-1)}{\cosh\mu}\bm{I}.
\end{equation}
The wave function renormalization $Z$ is fixed by the condition that the Goldstone boson is massless.
The implementation of this change in the algorithm is straightforward. Although theoretically cleaner we find that the
introduction of the parameter $Z$ has a negligible impact on the numerical solution: In the vicinity of the phase transition $(Z-1)$ is smaller than $10^{-4}$.
This is because the Goldstone boson is almost massless already and only a very small correction is needed.

\section{Results}\label{sec:results}\noindent
Just as in the real $\varphi^4$ theory \cite{Akerlund:2013} we find that EMFT predicts the location of the phase transition,
in this case the critical chemical potential $\mu_c$, with high accuracy. In Table~\ref{tab:muc} we summarize $\mu_c$ at zero temperature
for two values of $\eta$ at $\lambda=1$ for mean field theory, EMFT, Monte Carlo \cite{Gattringer:2012df} and complex Langevin \cite{Aarts:2009hn}.

\begin{table}
\caption{\label{tab:muc}Comparison of the critical chemical potential, $\mu_c(T=0)$ of four-dimensional complex $\varphi^4$-theory at $\lambda=1$
obtained by mean field theory, EMFT, Monte Carlo~\cite{Gattringer:2012df} and complex Langevin~\cite{Aarts:2009hn}.}
\begin{tabular}{l|c|c}
$\eta$ & $9.00$ & $7.44$\\\hline
Mean field theory & $1.12908$ & -\\
EMFT & $1.14582$ & $0.17202$\\
Monte Carlo & $1.146(1)$ & $0.170(1)$\\
Complex Langevin & $\approx 1.15$ & -\\
\end{tabular}
\end{table}

\begin{figure}[t]
\centering
\includegraphics[width=1\linewidth]{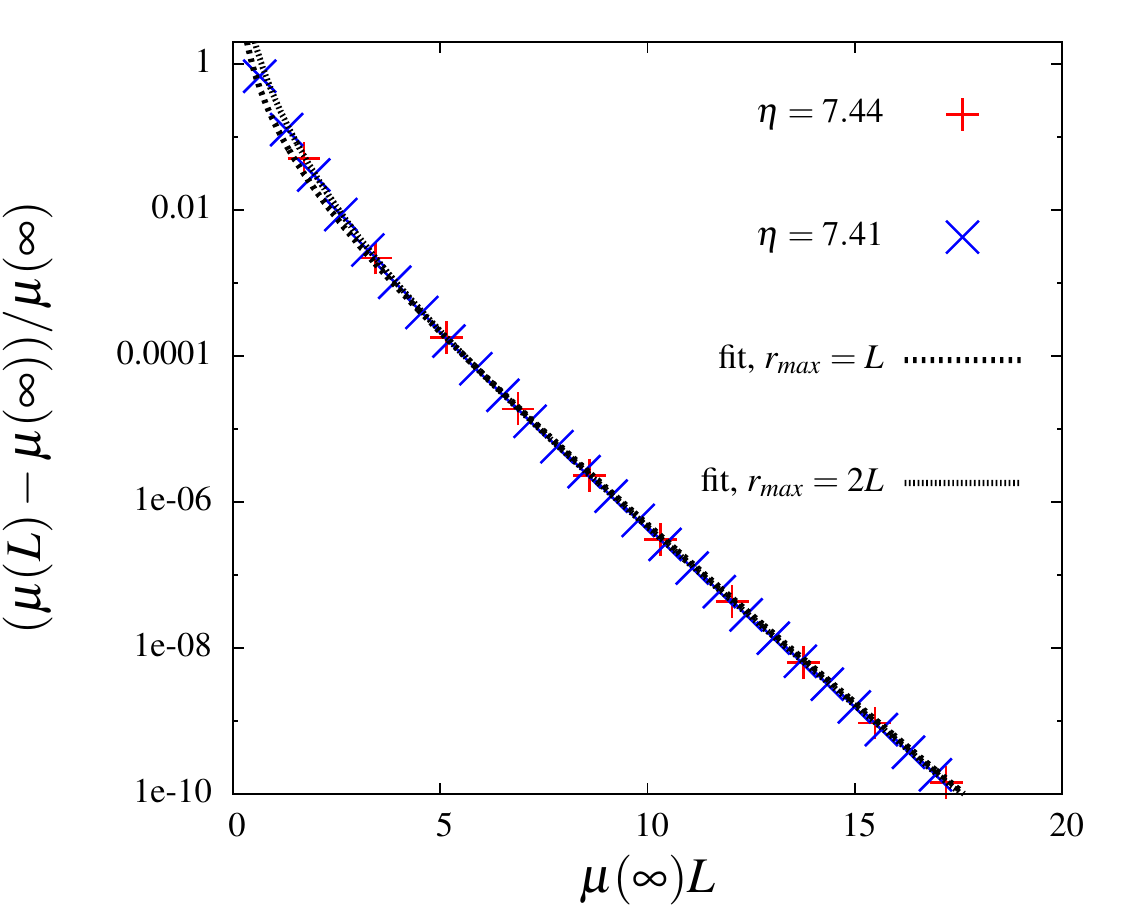}
\caption{The relative deviation of the critical chemical potential $\mu_c$ due to finite size effects as a function of the
spatial extent of the lattice, $L$, on an $L^3\times\infty$ lattice. We fit the amplitude of a sum of Yukawa potentials,
Eq.~\eqref{eq:yuk}, taking mirror particles up to the distance $r_\text{max}$ into account. In both cases the common amplitude
of the Yukawa potentials is the only free parameter. The mass is fixed to $\mu_c(\infty)$.}
  \label{fig:fv}
\end{figure}

We ultimately want to apply EMFT to models with nonzero temperature, but as a first test we will study the finite volume behavior
since it is more predictable. 
Let us vary the  spatial extent of the lattice and consider  the finite volume corrections to $\mu_c$. 
These arise since the particles
interact with their mirror images on the periodically continued lattice. Because the interaction is repulsive the mass will get
a positive correction at finite volume, $m(L)>m(\infty)$. The interaction is through particle exchange and hence
the potential is of the Yukawa type. The potential in four dimensions is given by
\be\label{eq:yuk}
V(r) \propto \frac{1}{r^2}(mr)K_1(mr),
\ee
where $K_1(x)$ is a modified Bessel function which decays exponentially for large arguments. The distance between two mirror
particles is $L = aN_s$. The decay is thus governed by $mL = (am_R)N_s$ which allows us to measure $am_R$ by considering
lattices of different sizes. Unless $mL$ is rather large it is important to consider particles which wind around the periodic
dimensions more than once. At criticality the correlation length diverges, i.e. the inverse propagator vanishes at $k=0$.
From the general form of the propagator (Eq.~\eqref{eq:free_gf}), 
\begin{equation}
  G^{-1}(k) = Z\left((am_R)^2 + 4\sum_\nu \sin\left(\frac{ak_\nu-ia\mu\delta_{\nu,t}}{2}\right)^2\right),
\end{equation}
we obtain $am_R = 2\sinh(a\mu_c/2)$, which reduces to $m_R = \mu_c$ in the continuum limit. 
In Fig.~\ref{fig:fv} we plot $(\mu_c(L)-\mu_c(\infty))/\mu_c(\infty)$
as a function of $\mu_c(\infty)L$ for two different values of $\eta$ together with the expected behavior, Eq.~\eqref{eq:yuk},
with the mass $m_R$ fixed to its infinite volume value $m(\infty)$. The results are largely independent of $\eta$, i.e. the finite
lattice spacing effects are negligible, and the mass in Eq.~\eqref{eq:yuk} is clearly given by $\mu_c(\infty)$.
We also see that at small volumes the mirror images at distances larger than $L$ start to play a role, but since we will work
directly in the thermodynamic limit in the following, this is of no concern to us.

\subsection{Finite temperature}
\begin{figure}[t]
\centering
\includegraphics[width=1\linewidth]{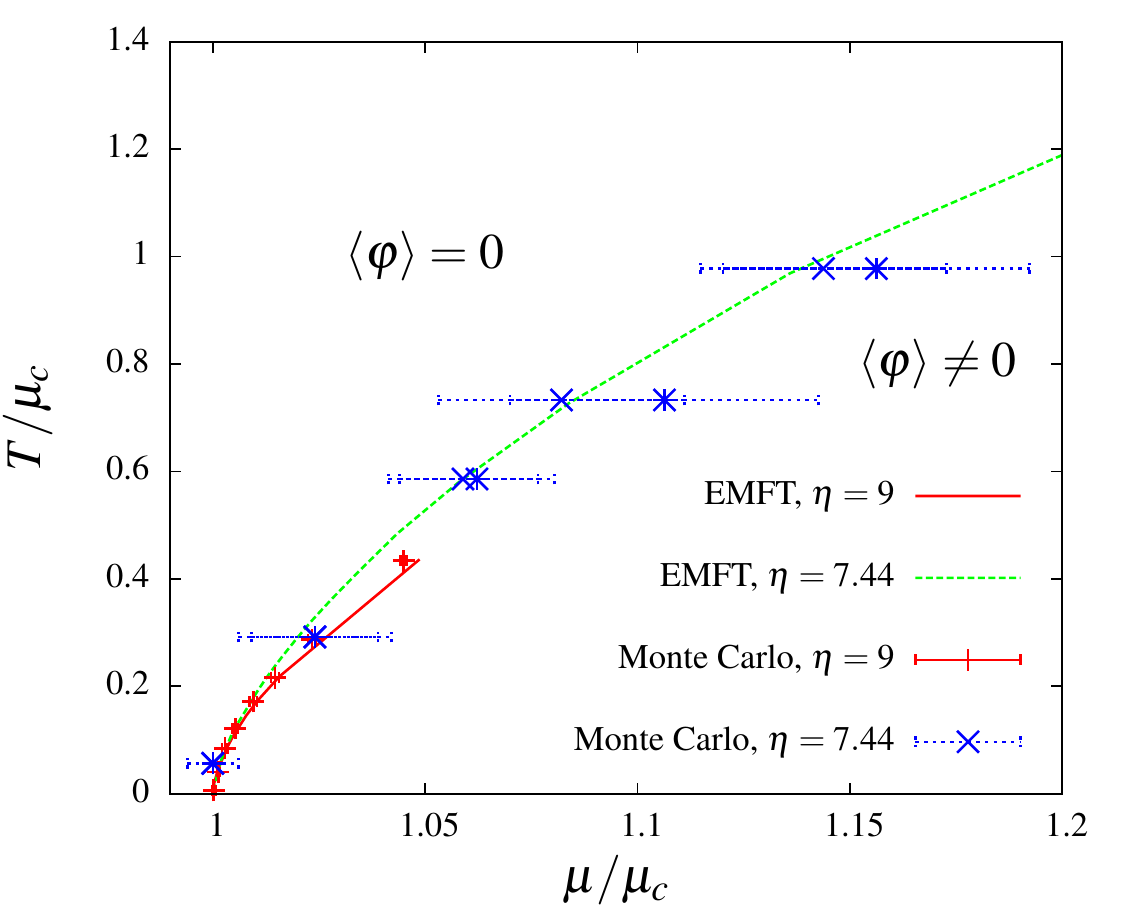}
\caption{$(T/\mu_c(T=0),\mu/\mu_c(T=0))$ phase diagram of complex $\varphi^4$ theory at $\lambda=1$ obtained by EMFT
and world line Monte Carlo (Gattringer and Kloiber \cite{Gattringer:2012df}). The two different blue symbols correspond to different observables used in determining the transition point, {\color{blue}\texttimes} for the variance of $\varphi$ and {\color{blue} \textasteriskcentered} for the density. We have used two values of $\eta$ and the results agree very well for both.}
  \label{fig:phasediag}
\end{figure}

One major advantage of EMFT over standard mean field theory is the access to finite temperature effects.
To turn on temperature we simply truncate the sum over $k_t$ in Eq.~\eqref{eq:sc_gf} at some finite value of $N_t$.
This lets us define a temperature in lattice units, $aT=N_t^{-1}$, or in units of the chemical potential,
$T/\mu=((a\mu)N_t)^{-1}$. By solving the self-consistency equations at different values of $N_t$ we can obtain all
observables as a function of the temperature at fixed lattice spacing. Our main result is the $(T/\mu_c,\mu/\mu_c)$
phase diagram which is shown in Fig.~\ref{fig:phasediag}. We have determined it for two lattice spacings, $\eta=9$ 
and $\eta=7.44$, to allow for a direct comparison with Monte Carlo results obtained by Gattringer and Kloiber \cite{Gattringer:2012df}.
In Ref.~\cite{Gattringer:2012df} the authors used a world-line formulation of the partition function, which has no sign
problem, and sampled the configuration space with a Monte Carlo algorithm. The agreement is excellent at all temperatures
and for both values of $\eta$.

\begin{figure}[t]
\centering
\includegraphics[width=1\linewidth]{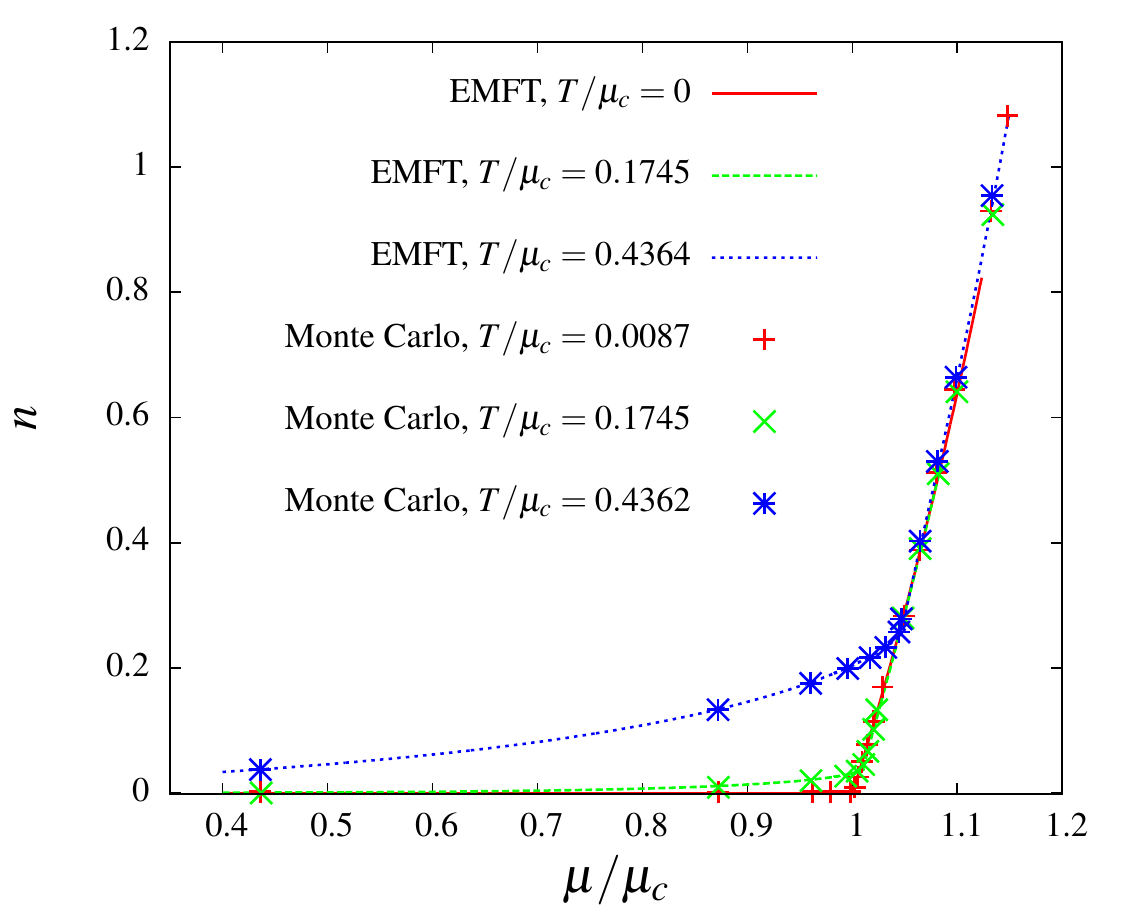}\\
\includegraphics[width=1\linewidth]{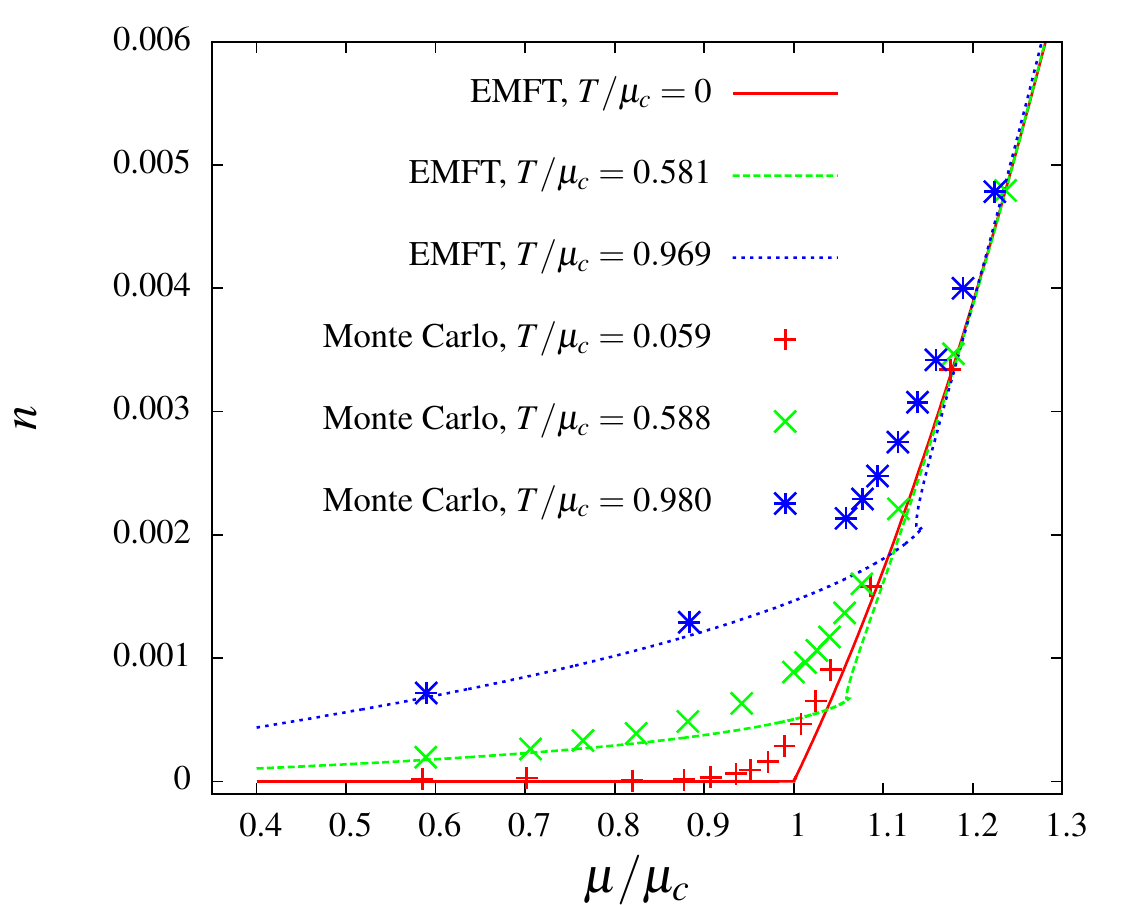}
\caption{The density $n$, Eq.~\eqref{eq:dens}, as a function on $\mu$ for a few different temperatures, $T/\mu_c \equiv 1/(N_t\mu_c(T=0))$,
at $\lambda=1$, $\eta=9$ (\emph{upper panel}) and $\lambda=1$, $\eta=7.44$ (\emph{lower panel}). The Monte Carlo data~\cite{Gattringer:2012df}
were obtained on a $N_s^3\times N_t$ lattice with $N_s=20$ for $\eta=9$ and $N_s=24$ for $\eta=7.44$. The small temperature
differences come from slightly different values of $\mu_c(T=0)$, see Tab.~\ref{tab:muc}.
The EMFT results are obtained in the thermodynamic limit, i.e. $N_s=\infty$.}
  \label{fig:dens}
\end{figure}

Also the EMFT estimate of the density as a function of $\mu$ at various temperatures agrees with the Monte Carlo
results to high accuracy. Again we compare to the Monte Carlo simulations in \cite{Gattringer:2012df} with $\lambda=1$, $\eta=9$ and $\lambda=1$, $\eta=7.44$. The result is shown in Fig.~\ref{fig:dens}. At $\eta=9$ (\emph{upper panel}) the finite
volume effects in the Monte Carlo data are small and the EMFT and Monte Carlo results agree almost perfectly. 
Since the nonzero temperature contribution to the density, Eq.~\eqref{eq:dens}, is closely related to the Green's
function at separation $a$, we conclude that EMFT is not restricted to predicting the local Green's function $G_{xx}$.
At $\eta=7.44$ (\emph{lower panel}), we are closer to the continuum limit, which means that the physical volume of the lattice is
smaller in the Monte Carlo simulation. This manifests itself as a rounding of the phase transition. This rounding is absent in
EMFT since the volume in these calculations is always infinite. Away from the transition the two methods agree very well also at the smaller value
of $\eta$.

\subsection{Dimensional reduction}

At nonzero temperature the theory is expected to undergo a dimensional reduction near the phase transition.
This is because the time extent of the lattice 
becomes much smaller than 
the correlation length.
In a lattice simulation of the full model it might be hard to see this happening for three reasons.
Firstly, it is expensive to increase the lattice volume, hence the time extent might not be a small enough
fraction of the spatial extent. Secondly, due to the Ginzburg criterion, the correlation length must not be
small compared to the time extent or the system will not realize dimensional reduction. Lastly, finite lattice
spacing corrections are of the form $a^2$ and might conceal the true critical behavior when large.
All this taken together provides a considerable challenge for Monte Carlo simulations.

\begin{figure}[t]
\centering
\includegraphics[width=1\linewidth]{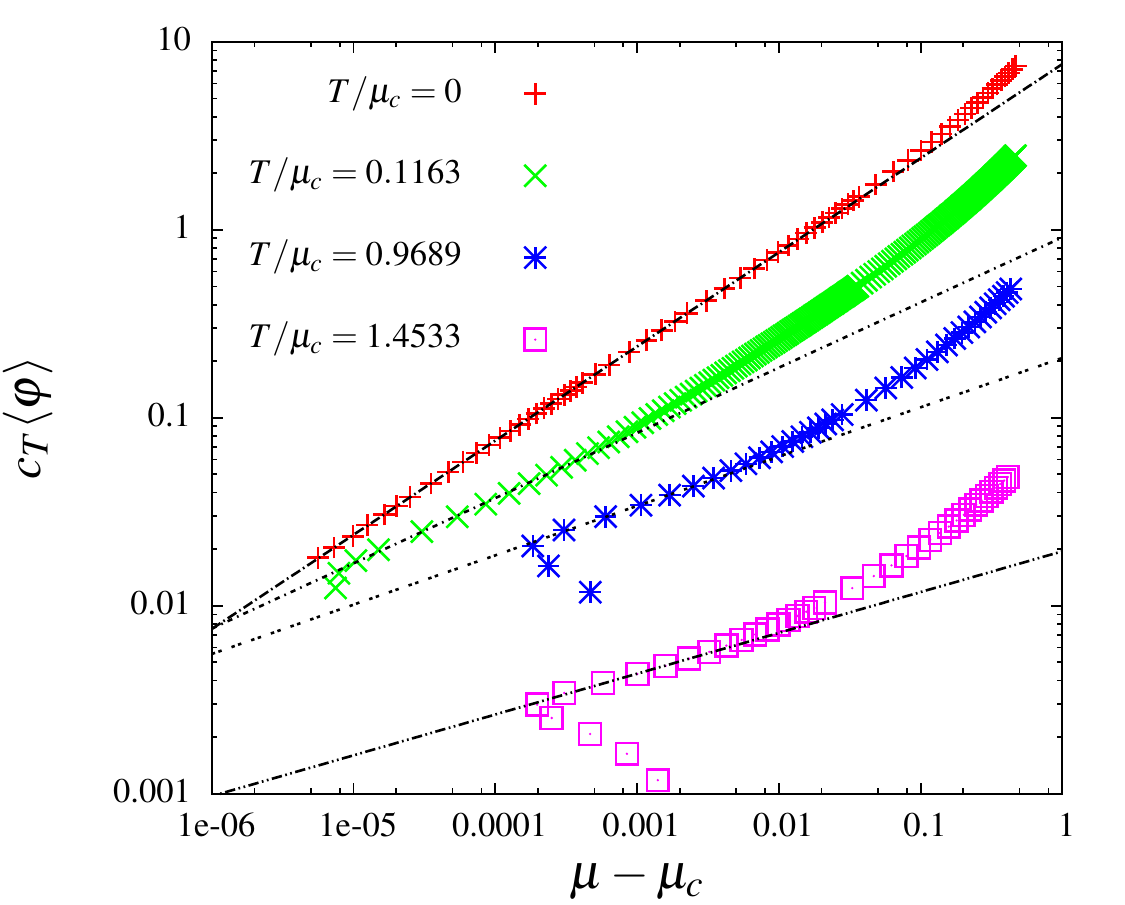}\\
\includegraphics[width=1\linewidth]{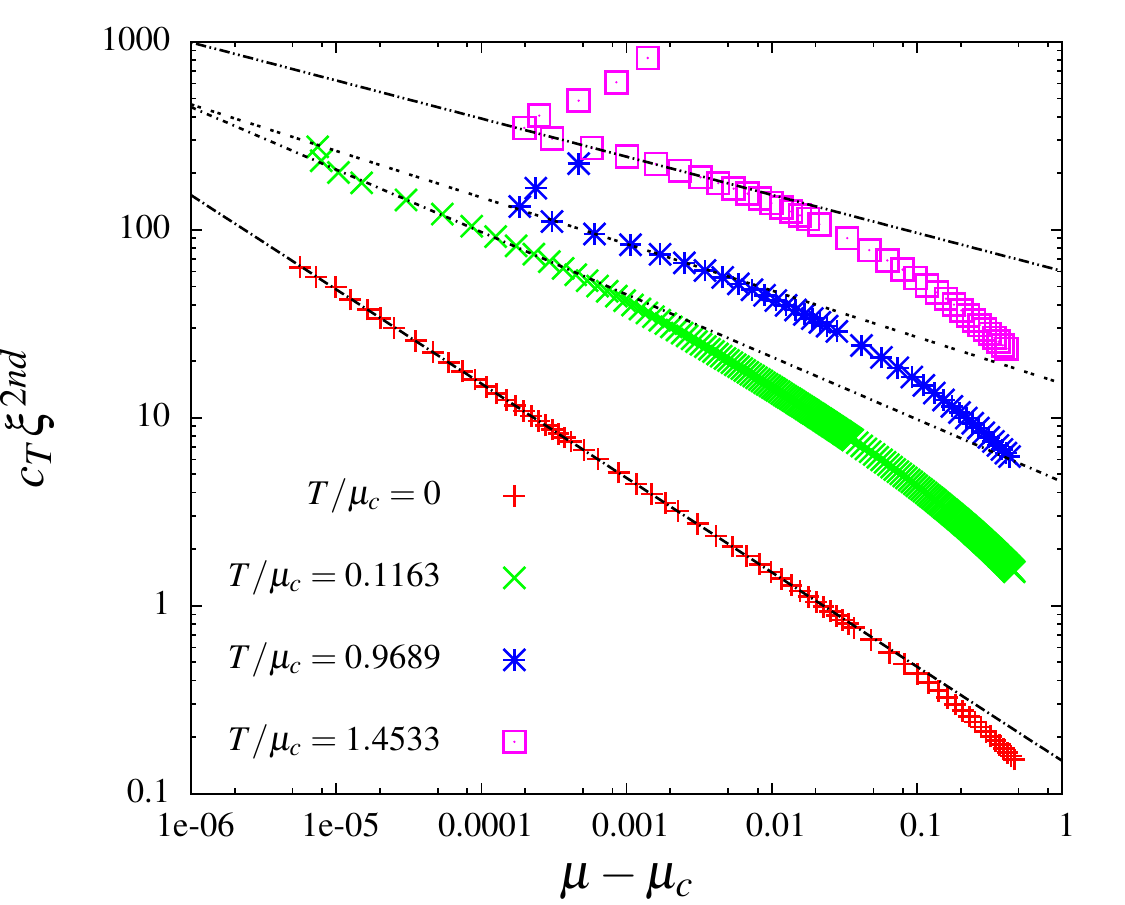}
\caption{The expectation value of the field, $\expv{\Re\varphi}$ (\emph{upper panel}), and the correlation length, $\xi$ (\emph{lower panel}),
as a function of the distance to the critical chemical potential for a few different temperatures at $\lambda=1$ and $\eta=7.44$ on a log-log scale.
We see how the power law changes from 0.5 at zero temperature to approximately $1/3$ at finite temperature for both observables.
To increase readability the curves have been multiplied by $c_T = 15,5,1$ and $0.1$, going from top to bottom.}
  \label{fig:dim_red}
\end{figure}

EMFT works best in the thermodynamic limit and does not suffer from critical slowing down close to the continuum
limit and can thus overcome all these problems. EMFT is, in other words, well
suited for an investigation 
of dimensional reduction. When $T>0$ we expect that the critical exponents change from mean field to those of the
three-dimensional $XY$-model universality class. Two critical exponents easily accessible to us are $\beta$ and $\nu$.
Fig.~\ref{fig:dim_red} shows the expectation value of the field and the correlation length as a function of $\mu-\mu_c$
for zero and nonzero temperatures. At zero temperature we find $\beta=\nu=0.50$. For finite temperature the power laws
change to approximately $1/3$ which is not the behavior expected for the $3d$ $XY$-model ($\beta\approx 0.33,\nu\approx 0.67$)
but rather what is expected close to a genuine first-order transition in three dimensions.

\subsection{First-order transition}
\begin{figure}[t]
\centering
\includegraphics[width=1\linewidth]{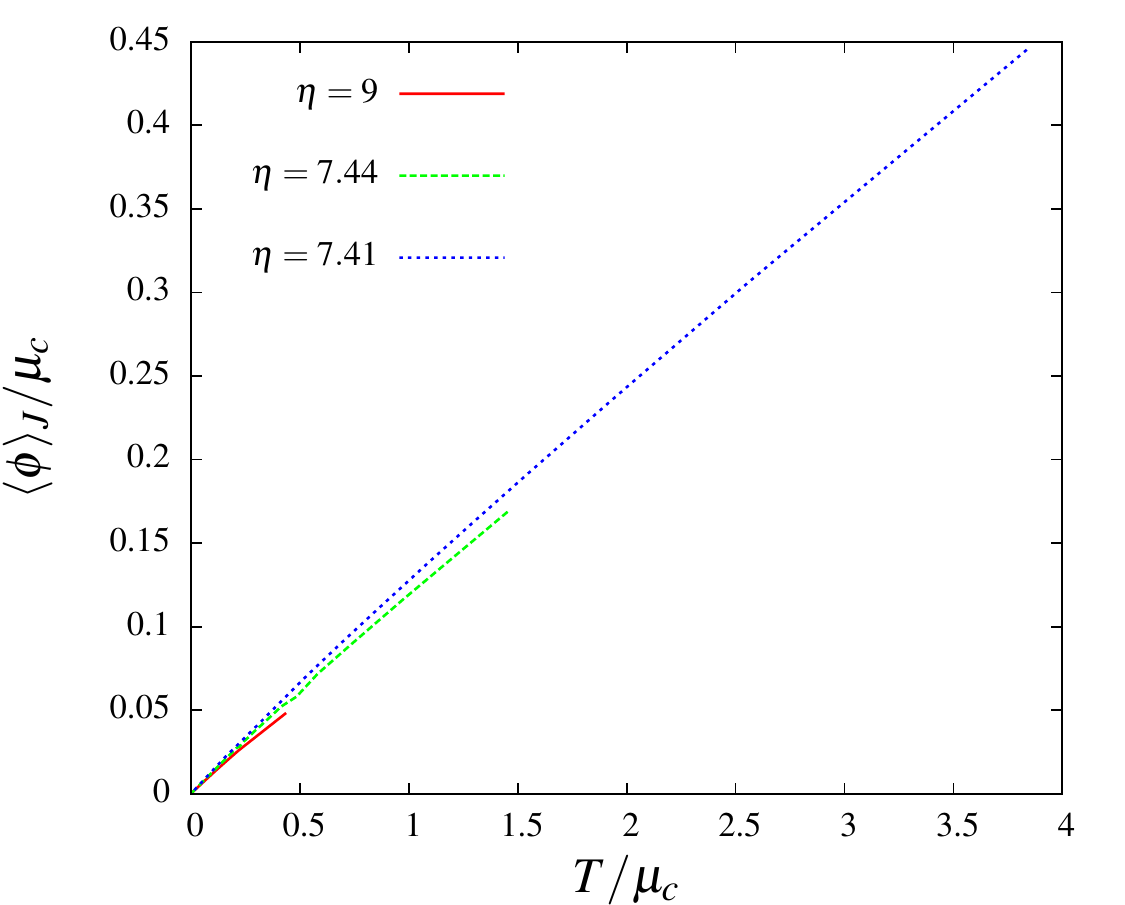}
\caption{The expectation value of the field at the chemical potential where $\partial\expv{\varphi}/\partial\mu=\infty$ as a function of the temperature,
both made dimensionless by division by $\mu_c(T=0)$.}
  \label{fig:jump}
\end{figure}

We have seen that the four-dimensional model dimensionally reduces as temperature is turned on, but the EMFT
incorrectly predicts a first order transition in this case. The strength of this first-order transition is
however quite weak, which can be seen from the value of the correlation length in Fig.~\ref{fig:dim_red}
(notice the shift of the curves). Although EMFT still produces quantitatively
good predictions of various observables such as the critical chemical potential and the density, this is
of course an undesired feature. It is interesting to quantify the strength of the first-order transition,
which can be done by determining how the jump in the expectation value depends on the temperature.
We define $\expv{\phi}_J$ to be the value of $\expv{\varphi}$ at the chemical potential where 
$\partial\expv{\varphi}/\partial\mu=\infty$ (cf. upper panel of Fig.~\ref{fig:dim_red}). In Fig.~\ref{fig:jump}
we plot $\expv{\phi}_J/\mu_c$ versus $T/\mu_c$. $\expv{\phi}_J$ grows slightly less than linearly in $T$
but seems to approach a linear behavior with a coefficient of about $0.14$ as we approach the continuum limit.

\section{Conclusions}\label{sec:conc}
We have demonstrated that EMFT works very well for complex $\varphi^4$ theory, a model 
for which conventional Monte Carlo simulations suffer from a sign problem. 
It works especially well in four dimensions at zero temperature where it correctly predicts a second-order
phase transition with mean field exponents and a quantitatively very accurate value of the critical chemical potential.
EMFT has also been shown to be a computationally cheap method for probing the system at finite temperature. Although it
incorrectly predicts a first-order transition due to dimensional reduction, the estimates of observables like the critical chemical potential and
the density agree very well with state of the art Monte Carlo simulations \cite{Gattringer:2012df}.
These properties make EMFT a potentially very useful tool for the study of the existence and whereabouts of phase transitions,
even though EMFT might have problems distinguishing a weak first-order transition from a second-order transition.
Due to its simplicity and low computational cost, it can serve as a complement and guide to more
sophisticated methods.

A natural and straightforward next step could be to study a model containing a multicomponent scalar, for example a
gaugeless SU(2) Higgs model or Higgs-Yukawa models. An even more interesting extension would be to include the
gauge field and study for example a U(1) Higgs model. Since the plaquette, the smallest gauge-invariant object, is an
extended object we would have to generalize the method to work with a cluster of live sites. Such an extension is
interesting in its own right since it would allow for a self-consistent determination of momentum-dependent observables.
By taking larger and larger clusters of live sites it is also possible to systematically approach the full model again.
That could be useful for assessing the accuracy of the method in a case where an \emph{ab initio} calculation is not possible
or has not been done. 

\section*{Acknowledgments}
We thank Christof Gattringer and Thomas Kloiber for discussions and correspondence.

\appendix
\section{$k$-integrated Green's functions}\label{app:kint}
Our goal is to efficiently calculate the local Green's function from the Green's function in momentum space.
This is equivalent to integrating it over all momenta,
\be
\bm{G}_{xx} = \int\!\frac{\rd^dk}{(2\pi)^d}\,\widetilde{\bm{G}}(k).
\ee
The main complication is that we only know $\widetilde{\bm{G}}^{-1}(k)$ explicitly. Let us consider the general
case where we have $N$ real or $N/2$ complex fields. In this case the free Green's functions form a diagonal 
$N\times N$ matrix and the EMFT Green's functions form a full matrix,
\begin{align}
\widetilde{\bm{G}}(k) &= \left[\bm{G}^{-1}_\text{EMFT}+\bm{\Delta} - 2\sum_{\nu=1}^d\cos(k_\nu-i\mu\delta_{\nu,t})\bm{I}\right]^{-1}\nonumber\\
&\equiv\left[\bm{A}-\epsilon(k,\mu)\bm{I}\right]^{-1},
\end{align}
with the kinetic part
\be
\epsilon(k,\mu) = 2\sum_{\nu=1}^d\cos(k_\nu-i\mu\delta_{\nu,t}).
\ee
The self-energy matrix $\bm{A}$ can be found by inverting the measured $\bm{G}_\text{EMFT}$. Now, when $N=1$, we can rewrite this
in a form which allows for an analytic integration of the $d$ components of $k$,
\be
\frac{1}{a-\epsilon(k,\mu)} = \int_0^\infty\!\!\rd\tau\,e^{-a\tau}\prod_{\nu=1}^d e^{2\tau\cos(k_\nu-i\mu\delta_{\nu,t})}.
\ee
We can integrate over $k$ by using an integral representation of the modified Bessel function of first order, $I_0(x)$,
\be
I_0(x) = \int_{-\pi}^\pi\frac{\rd k}{2\pi}\,e^{x\cos(k+z)}.\label{eq:bessel}
\ee
Note that the (complex) constant $z$ is irrelevant. The final result reads
\be
\int\frac{\rd^dk}{(2\pi)^d}\,\frac{1}{a-\epsilon(k,\mu)}=\int_0^\infty\rd\tau\,e^{-a\tau}I_0^d(2\tau).\label{eq:k_int}
\ee
To study finite volume (temperature) we simply replace the relevant Bessel functions with what is obtained when 
the integral in Eq.~\eqref{eq:bessel} is replaced by a discrete sum.

We will now show that $\widetilde{\bm{G}}(k)$ can be written as a sum of such integrable terms for any value of $N$.
Since $\widetilde{\bm{G}}^{-1}(k)$ is symmetric and the $k$ dependence is only on the diagonal, $\widetilde{G}^{-1}$ is diagonalized by
a $k$-independent orthogonal matrix $\bm{U}$ which also diagonalizes $\widetilde{\bm{G}}(k)$.
The eigenvalues which make up the diagonal $\widetilde{\bm{D}}(k) = \bm{U}^\intercal\widetilde{\bm{G}}(k)\bm{U}$
are given by $\left(\lambda_i-\epsilon(k,\mu)\right)^{-1}$ where \{$\lambda_i\}_{i=1}^N$ are the $N$ eigenvalues of $\bm{A}$.
Using the $k$ independence of $\bm{U}$ we just have to integrate the elements of $\widetilde{\bm{D}}(k)$,
which are all integrals of the form of Eq.~\eqref{eq:k_int}. The matrix elements of $\bm{G}(0)$ are then trivially
recovered by applying $\bm{U}$. Explicitly they are given by
\begin{equation}\label{eq:inv_sum_int}
  \left(\bm{G}_{xx}\right)_{ij} = \sum_{k=1}^N\bm{U}_{ik}\bm{U}_{jk}\int_0^\infty\!\rd\tau\,e^{-\lambda_k\tau}I_0^d(2\tau).
\end{equation}
So, instead of performing one complicated $d$-dimensional integral for each matrix element, we can diagonalize the
matrix and compute $N$ one-dimensional integrals.

\section{Finite temperature contributions to the density}\label{app:dens}
In Sec.~\ref{sec:emft} we derived a formula for the density, Eq.~\eqref{eq:dens}. We will here show that the
second part vanishes at zero temperature and gives a positive contribution for nonzero temperatures. We assume
here that $\mu\geq 0$ but note that the density is odd in $\mu$.
We have to deal with the two integrals
\begin{align}
I_R&  \equiv\sinh\mu \int\frac{\rd^d k}{(2\pi)^d}\Re[\expv{\varphi^*(k)\varphi(k)}_c]\cos(k_t),\\
I_I&  \equiv\cosh\mu \int\frac{\rd^d k}{(2\pi)^d}\Im[\expv{\varphi^*(k)\varphi(k)}_c]\sin(k_t),
\end{align}
where the correlator is the diagonal element of $\widetilde{\bm{G}}(k)$. To decouple $k_t$ from the other momenta
we use the same trick as in Appendix~\ref{app:kint}.
Considering only the integral over $k_t$ we have
\begin{align}
I_R &\propto \sinh\mu\int\frac{\rd k_t}{2\pi} \Re\left[\exp\left(2\tau\cos(k_t-i\mu)\right)\right]\cos(k_t),\\
I_I &\propto \cosh\mu\int\frac{\rd k_t}{2\pi} \Im\left[\exp\left(2\tau\cos(k_t-i\mu)\right)\right]\sin(k_t),
\end{align}
where $\tau$ is an auxiliary integration variable. Noting that everything not depending on $k_t$ is the same
for the two terms we find after some algebra that the difference is proportional to
\begin{align}
&\Re\left[\int\frac{\rd k_t}{2\pi}\exp(2\tau\cos(k_t-i\mu))\big(\exp(i(k_t-i\mu))\right.\nonumber\\
&\left.\phantom{\Re[\int\frac{\rd k_t}{2\pi}}-\exp(-i(k_t-i\mu))\big)\right].
\end{align}
This expression can be further simplified using the modified Bessel function identity
\be
\exp(z\cos(w)) = \sum_{l=-\infty}^\infty I_l(z)\exp(iwl).
\ee
The integrand is just a sum of weighted exponentials, $\exp(ik_tn)$ for integer $n$, and the $k_t$ integral
is nonvanishing only when $n=0$. This selects $I_{-1}(2\tau)$ and $I_1(2\tau)$, which are identical for
real arguments, hence the difference vanishes. If we consider a nonzero temperature the momentum can only
take discrete values, $k_t = \tfrac{2\pi}{N_t}n,\, n\in\{0,\ldots,N_t-1\}$,
and the sum over $n$ yields a nonzero contribution when $l+1 = \pm m N_t$. Combining the two we find 
\begin{align}
I_R+I_I &\propto \sum_{l=1}^\infty\left(I_{N_tl-1}(2\tau)-I_{N_tl+1}(2\tau)\right)\sinh(\mu N_tl) \nonumber\\
&= \frac{N_t}{\tau}\sum_{l=1}^\infty l I_{N_tl}(2\tau)\sinh(\mu N_tl),
\end{align}
which is positive and goes to zero as $N_t\to\infty$.
\end{document}